\documentclass[conference]{IEEEtran}
\ifCLASSINFOpdf
\else
\fi
\usepackage[numbers]{natbib}
\usepackage{epsfig}
\usepackage{epstopdf}
\usepackage[T1]{fontenc}
\usepackage{amssymb}
\usepackage{amsmath}
\usepackage{amsfonts}
\usepackage{verbatim}
\usepackage{graphicx}
\usepackage{hyperref}
\usepackage[usenames,dvipsnames]{xcolor}
\usepackage{algorithmic}[5]
\usepackage{algorithm}
\usepackage{array}
\usepackage{multirow}
\usepackage{caption}
\usepackage{subcaption}
\usepackage{multicol}
\usepackage{pgfplots}
\usepackage{lipsum}
\usepackage{colortbl}
\usepackage{pifont}
\begin{document}
%
\title{The Blockchain Based Auditor  on  Secret key Life Cycle in Reconfigurable Platform}

%
\author{\IEEEauthorblockN{Rourab Paul}
\IEEEauthorblockA{
Computer Science \& Engineering, \\ Siksha 'O' Anusandhan (Deemed \\to be
University), Odisha, India\\
rourabpaul@soa.ac.in} 
\and
\IEEEauthorblockN{Nimisha Ghosh}
\IEEEauthorblockA{
Computer Science \& IT, \\ Siksha 'O' Anusandhan (Deemed \\to be
University), Odisha, India\\
nimishaghosh@soa.ac.in}
\and
\IEEEauthorblockN{Amlan Chakrabarti}
\IEEEauthorblockA{
School of IT, \\ University of Calcutta\\
Kolkata, India\\
achakra12@yahoo.com}
\and
\IEEEauthorblockN{Prasant Mahapatra}
\IEEEauthorblockA{
Dept. of Computer Science \\University of California, USA\\
pmohapatra@ucdavis.edu}
}


%


\maketitle

\begin{abstract}
The growing sophistication of cyber attacks, vulnerabilities in high computing systems and increasing dependency on cryptography to protect our digital data make it more important to keep secret keys safe and secure. Few major issues on secret keys like incorrect use of keys, inappropriate storage of keys, inadequate protection of keys, insecure movement of keys, lack of audit logging, insider threats and non-destruction of keys can compromise the whole security system dangerously. In this article, we have proposed and implemented an isolated secret key memory which can log life cycle of secret keys cryptographically using  blockchain (BC) technology. We have also implemented a special custom bus interconnect which receives custom crypto instruction from Processing Element (PE). During the execution of crypto instructions, the architecture assures that secret key will never come in the processor area and the movement of secret keys to various crypto core is recored cryptographically after the proper authentication process controlled by proposed hardware based BC. To the best of our knowledge, this is the first work which uses  blockchain based solution to address the issues of the life cycle of the secret keys in hardware platform. The additional cost of resource usage and timing complexity we spent to implement the proposed idea is very nominal. We have used Xilinx Vivado EDA tool and Artix 7 FPGA board.
\end{abstract}
\begin{IEEEkeywords}
Blockchain, FPGA, Key Memory, Secret Key Life Cycle
\end{IEEEkeywords}
\vspace{-10pt}

%
\IEEEpeerreviewmaketitle

\section{Introduction}
The security of a crypto system depends on three primary keys. \textbf{Symmetric Key}: is used to encrypt bulk data in symmetric key algorithms like AES, TDES, DES etc. \textbf{Private Keys}: of public-private key pair used in Asymmetric Key Cryptography such as RSA, Diffie Helmen etc. Private key is used for signature generation and key exchange process. \textbf{Hash Key}: is used to check integrity and authenticity  of transactions and data with algorithms like SHA-3. Increasing volume of secret key and data protected by those keys makes the cryptographic key management relevant in current research trend \cite{cryptomathic}. There are several threats causes compromised key. In this article we have discussed about few major threats.
\subsection{Storage }
In many sense, processor based architectures are flexible but it can be exposed to software threats like cache attack\cite{coldboot:sim}, bus snooping \cite{bus_en}, memory disclosure attack  \cite{swmem:dis:guan} etc. hence the storage of secret keys should be separated form software area and it should be isolated from any physical connection of processor.  
\subsection{Insecure Movement}
A security processor allows key movements between several cores like Key Memory, RNG, Hash, Symmetric Key and Asymmetric Key core. All these key movement to the key memory and from the key memory should be authenticated and secured. These transactions of key mainly suffers two issues. \textbf{A} The buses used for these movements can be snooped by software attacks. \textbf{B} The request of key can be issued from some compromised IP placed inside the architecture. To overcome attack-A the most effective solutions are bus encryption \cite{bus_en}, or partition of software area from these buses \cite{aes:mccp}. Attack-B can be prevented by verifying the signature of the requestee IP. To the best of our knowledge we did not find any similar solution in hardware platform.
\subsection{Non-destruction}
Once the signature of requestee IP is verified, the key should be released to the proper destination. Except the pre-master key, other keys should be destroyed securely. This removal of key  should be logged cryptographically and should non-traceable.
\subsection{Audit Logging}
The secret keys of a security processor is the most sensitive data. The creation, deletion and the movement of secret key should be logged and audited cryptographically otherwise it will be difficult to identify a compromise for forensic investigation.  
\subsection{Incorrect use of keys}
The generated master keys are for specific purposes. Hash core can not use encryption key and vice versa. If the keys are used for something else, the proper protection action should be taken.
The main contribution of this article are stated below:
\begin{itemize}
\item This article proposes a security processor architecture which is partitioned in three separated areas such as processor area, crypto area, and confidential area. This architecture assures that secret key memory is isolated form processor area.
\item The architecture proposes a blockchain based auditor to monitor secret key lifecycles. The architecture prevents insecure movement, unauthorized key request, non-destruction issued and incorrect use of keys.  
\item The proposed security processor can execute 21 custom crypto instructions as shown in Table \ref{ins}.
\end{itemize}
\begin{figure}[!htb]
\centering
\vspace{-10pt}
\includegraphics[scale=0.4]{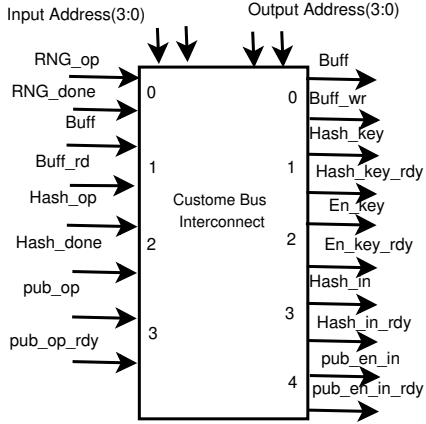}
\caption{Custom Bus Interconnect}
\vspace{-10pt}
\label{fig:cbi}
\end{figure}

\begin{figure}[!htb]
\centering
\vspace{-10pt}
\includegraphics[scale=0.22]{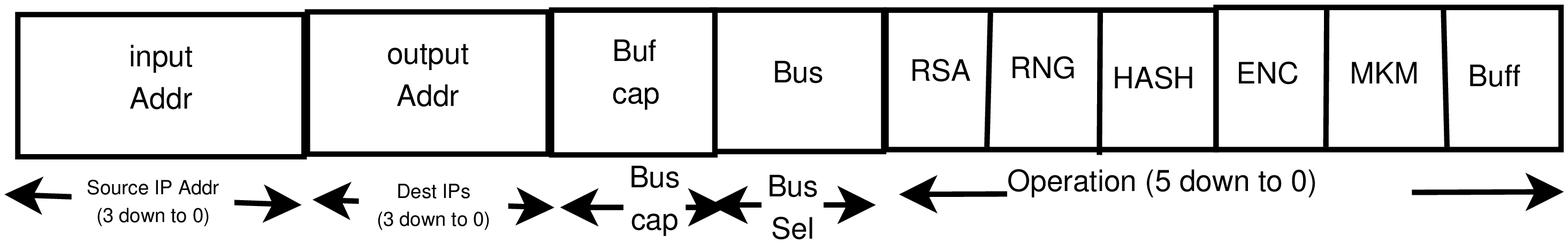}
\vspace{-6pt}
\caption{Control Word Register}
\vspace{-8pt}
\label{fig:cwd}
\vspace{-12pt}
\end{figure}
\section{Architecture}
In this architecture we have partitioned three areas such as \textbf{(i)} processor area, \textbf{(II)} crypto area and \textbf{(iii)}confidential area. The whole architectural details is stated in \cite{robupaul}. The partitioned and isolated confidential area to store secret keys assures that the secret key never come in the processor area because the processor area is very much vulnerable for various software threats. The main contribution of the proposed article is to add a special private blockchain to audit the movement of secret keys. Each secret key movement is authenticated and registered cryptographically. To adopt blockchain, we have designed the $data~path~controller$ (DPC), $custom~bus~interconnect$ (CBI), $buffer$ and a $Signature~Checker$ (SC) core. The architecture can execute 21 crypto instructions divided in 5 categories as shown in Table \ref{ins}. All these proposed instructions are executed by $DPC$ and  $CBI$. As instructed by the TLS white paper \cite{rfc:tls}, the architecture processes instructions for $pre-master~key~generation$, $master~key~generation$, $hash$ and $encryption$ process. For the proposed blockchain process we have created a new category named as $common~purpose$. The details of other instructions are not described because of page restriction. It is to be noted that blockchain data is written in main memory which can be accessed by the PE.
\begin{figure}[!htb]
\centering
\includegraphics[scale=0.25]{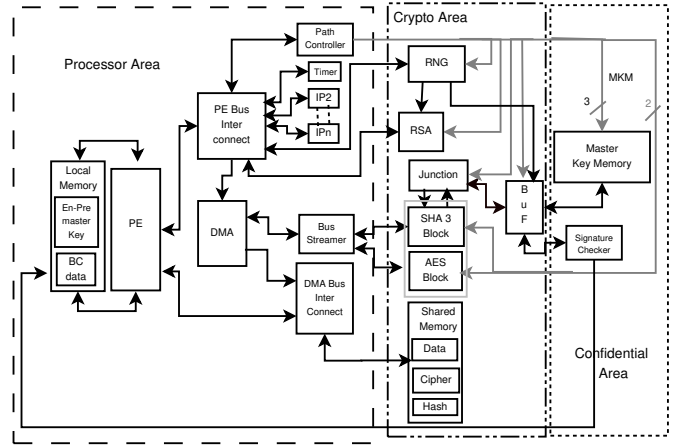}
\caption{System Architecture}
\vspace{-15pt}
\label{fig:sys}
\end{figure}
\subsection{Custom Bus Interconnect}
The proposed $custom~bus~interconnect$ (CBI) is a junction through which various crypto cores and the $buffer$ which is created as gateway of $Master~Key~Memory$ (MKM)  can communicate. Various paths of $crypto~cores$ and $buffer$ are controlled by 4 bits input and 4 bit output addresses. Input-output address pins are controlled by $data~path~controller$ (DPC) IP. In our current version of design, we have 4 inputs coming from $buffer$, output of $hash$, output of $RNG$ and output of $RSA$.  The 5 outputs of  $CBI$ are connected with $buffer$, key of $AES$, key of $hash$ and key of $RSA$ block. The synchronization signals of input cores and output cores are also taken care of by $CBI$. The fig. \ref{fig:cbi} shows the synchronization signals such as $RNG\_done$, $Buff\_rd$, $Hash\_done$, $Buff\_rdy$, $Hash\_key\_rdy$ and $En\_key\_rdy$. The $CBI$ is combination of MUX and DE-MUX where MUX's output is connected with DEMUX's input. The input address and output address pins are the selecting inputs of MUX and DE-MUX respectively.
\subsection{Data Path Controller}
The proposed $data~path~controller$ (DPC) is a slave IP of $PE$. The $PE$ can send $21$ instructions by Application Peripheral Interface (API) call for various crypto operations of TLS protocol. The $data~path~controller$ maintains a control word register to execute the 21 instructions as shown in Table \ref{ins}. The PE can communicate  $DPC$ by AXI bus through $PE~bus~interconnect$. In the current version of hardware, $DPC$ has 16 bits control word to control various crypto cores placed in the crypto area. The data path and binaries for all proposed instructions are shown in \ref{ins}. As shown in fig. \ref{fig:cwd}, the 11th bit is set to logic $1$ to enable the $custom~bus~interconnect$, otherwise it will be logic $0$ to disable the said interconnect. The 16th to 13th bits are used to select the input address. If the $input~address$ is $0000$, $RNG$ will be selected. It will be $0001$, $0010$ and $0011$ to select $Buff$, $Hash$ and $PubEn$ core respectively. The 12th and 9th bit is to address outputs. The $output~address$ will be $0000$, $0001$, $0010$, $0011$ and $0100$ to select the $Buff$, $Hash\_Key$, $En\_key$, $Hash\_In$ and $Pub\_en\_in$ core. The 6th to 1st bit of $control~word~register$ (CWR) are used to enable different hardware blocks such as $RSA$, $RNG$, $Hash$, $Enc$, $MKM$ and $Buff$. The PE directly can write to CWR to control the data path of the hardwares placed in the $crypto~area$.
\subsection{Buffer}
The proposed buffer is a gateway to write or read keys to the $MKM$. This buffer can accommodate $data$, $timestamp$ from $timer$ IP, $pre-hash$ from $Signature~Checker$, read/write operation, $signature$ of data and the system status. The system status comes from the enable ports and $ready$ ports of all the existing IPs including crypto cores. Instruction-1 writes $384$ bit random number from $RNG$ to the data portion of $Buffer$. Similarly instruction 9, 17 and 18 can write 512 bit data from KECCAK and 1024 bits from RSA respectively.
\subsection{Signature Checker}
The $Signature~Checker$ (SC) consists two main parts such as KECCAK hash and RSA block to verify signature of the IPs requested to read or write to $MKM$. This IP receives the signature from $buffer$. The signature is the encrypted hash of data. The $SC$ first decrypt it by the RSA block with the public key of requestee IP. The requestee IP is addressed by the source IP address available in $CWR$. After that SC gets the hash of data. The data is already available in the $buffer$. If the hash of $buffer$ data is matched with the decrypted data by RSA then the requested transaction will be granted by $SC$.
\begin{table}
	\setlength{\tabcolsep}{2.5pt}
	\caption{Comparison Results}
	\label{res}
	\resizebox{7cm}{!}{
		\begin{tabular}{|c|c|c|c|c|c|c|}\hline
		\#&\shortstack{key \\storage} &\shortstack{Incorrect\\use}& \shortstack{Insecure \\movement}&\shortstack{Destruction}&\shortstack{Audit\\log}\\\hline		
\shortstack{Propo\\sed}&\shortstack{key\\ memory} &$\times$&$\times$&$\times$&\checkmark\\\hline		
\shortstack{\cite{aes:mccp}\\2011}&\shortstack{key\\ memory} &\checkmark&\checkmark&\checkmark&$\times$\\\hline
\shortstack{\cite{hcrypt}\\2010}&\shortstack{key \\memory} &\checkmark&\checkmark&\checkmark&$\times$\\\hline
\shortstack{\cite{robupaul}\\2018}&\shortstack{key \\memory} &\checkmark&\checkmark&\checkmark&$\times$\\\hline
\shortstack{\cite{aws}\\2019}&\shortstack{main \\memory} &\checkmark&\checkmark&$\times$&$\checkmark$\\\hline
\shortstack{\cite{bus_en}\\2009}&\shortstack{main \\memory} &\checkmark&$\times$&\checkmark&$\times$\\\hline
\shortstack{\cite{coldboot:sim}\\2015}&\shortstack{cache \\memory} &$\times$&$\checkmark$&\checkmark&$\times$\\\hline
\shortstack{\cite{dis}\\2007}&\shortstack{main \\memory} &$\checkmark$&$\checkmark$&\checkmark&$\times$\\\hline
		\end{tabular}
	}
	\\ \checkmark=possible, $\times$=not possible\\
	\vspace{-20pt}
\end{table}
\begin{table}
	\setlength{\tabcolsep}{2.5pt}
	\vspace{-0.9em}    
	\caption{Instructions Implemented in Proposed Architecture}
	\label{ins}
	\resizebox{9cm}{!}{
		\begin{tabular}{|c|c|c|c|c|c|c|}\hline
		&\#&Instruction &Flow& Function&\shortstack{CWR\\(hex)}&\shortstack{Bus\\Inter \\Connect}\\\hline
		
		&1&\shortstack{PE Send \\Re-Seed }&PE to RNG &\shortstack{Send new seed to RNG from PE.R-\\ -NG needs seed to generate rand. no.}&0010&AXI\\\cline{2-7}
	     \rotatebox{90} {\hspace{-13pt} RNG}  & 2& Gen RND &RNG to Buff&\shortstack{Enable RNG to Generate Random \\Number, and write RND in data\\ portion of Buffer}&0050&Custom\\\cline{2-7}
	         \rotatebox{90} {Master by}   &3&\shortstack{RNG Write \\Block Gen} &to Buff&\shortstack{Block Generation for Write Operation\\ on MKM from RNG. Buff reads, $time$\\$stamp$ from $timer$ IP, $prehash$ from\\ $Signature~Checker$, write operation \\ the system status, source \& dest. IP ID}&0091&Custom\\\cline{2-7}
		  \rotatebox{90} { Pre}  &4&\shortstack{Re-Key\\ RSA } & PE to RSA&\shortstack{PE writes pub key of server to RSA}&0020&AXI\\\cline{2-7}
		&5&\shortstack{PE Get\\ En-RNG }& RSA to PE&\shortstack{Encrypt RND using server pub key}&XXXX&AXI\\\hline
		
		&6&\shortstack{PE Send Rands\\ in Hash}& \shortstack{PE to Hash\\ by DMA}&\shortstack{PE sends server random number \\and client random number to Hash}&XXXX&AXI\\\cline{2-7}
		  \rotatebox{90} {Master Key}  &7&\shortstack{Hash Read \\Block Gen }&to Buff&\shortstack{Block Generation for Read Operation\\ from MKM by Hash. Buff reads , $time$\\$stamp$ from $timer$ IP, $prehash$ from\\ $Signature~Checker$, read operation, \\ the system status, source \& dest. IP ID}&11C1&Custom\\\cline{2-7}
		&8&\shortstack{Get M-Key\\ Hash }& Buff to Hash&\shortstack{read master key to generate four keys}&1149&Custom\\\cline{2-7}
		&9&Gen Keys&Hash to Buff & \shortstack{Write keys to Buff}&2049&Custom\\\cline{2-7}
		&10&\shortstack{Hash Write \\Block Gen} & Hash to Buff&\shortstack{Block Generation for write Operation\\ to MKM}&20C9&Custom\\\hline
			  \rotatebox{90} {\hspace{-20pt} Encryption} &11&\shortstack{En Read \\Block Gen} &  to Buff&\shortstack{Block Generation for Read Operation\\ from MKM by En. Buff reads , $time$\\$stamp$ from $timer$ IP, $prehash$ from\\ $Signature~Checker$, read operation, \\ the system status, source \& dest. IP ID}&12C1&Custom\\\cline{2-7}
		&12&\shortstack{Get EN Key \\from Buff}&Buff to En&\shortstack{Read En. key from buffer}&1245&DMA\\\cline{2-7}
		&13&GEN EN&Shared to Shared&\shortstack{AES block reads plaintext form SM \\\& write cypher to SM  using DMA}&XXXX&Custom\\\hline
		&14&\shortstack{Hash Read \\Block\ Gen} &  to Buff&\shortstack{Same as Instruction 7}&11C1&Custom\\\cline{2-7}
		
		&15&\shortstack{Get Key\\ Hash}&Buff to Hash&Same as Instruction 8&1149&Custom\\\cline{2-7}
		  \rotatebox{90} {\hspace{15pt}Hash}&16&GEN Hash&Shared to Shared&\shortstack{KECCAK block reads plaintext form SM \\\& write digest to SM  using DMA}&XXX&DMA\\\hline
		
		&17&\shortstack{Hash of\\ Buff}  & load Buff to Hash &\shortstack{1st step of signature: load \\ Buff data to Hash input}&1341&Custom\\ \cline{2-7}
		&18&\shortstack{Hash \\to Buff } & Hash Op to Buff&\shortstack{2nd step of signature: load hash \\output  to buffer}&2049&Custom\\ \cline{2-7}
		 \rotatebox{90} {\hspace{-3pt}e}&19&\shortstack{PubEn \\of Buff} & Buff to PubEn&\shortstack{3rd step of signature: load buffer \\ in PubEn input}&1461&Custom\\ \cline{2-7}
		  \rotatebox{90} {Signatur}&20&\shortstack{PubEn \\to Buff }& \shortstack{Output of public \\key encrypted\\ data to buffer}&\shortstack{$4^{th}$ step of signature: load encrypted da-\\ta of PubEn to buffer which is the signa-\\ture of buffer data in 1st step}&3061&Custom\\ \cline{2-7}
		&21&Verify Sig &Buff to MKM&\shortstack{if signature matched\\ buffer data sent to MKM}&1XX3&Custom\\\hline

		\end{tabular}
	}
\end{table}
	\vspace{-20pt}
\section{Implementation \& Results}
\begin{figure}[!htb]
\centering
\vspace{-15pt}
\includegraphics[scale=0.36]{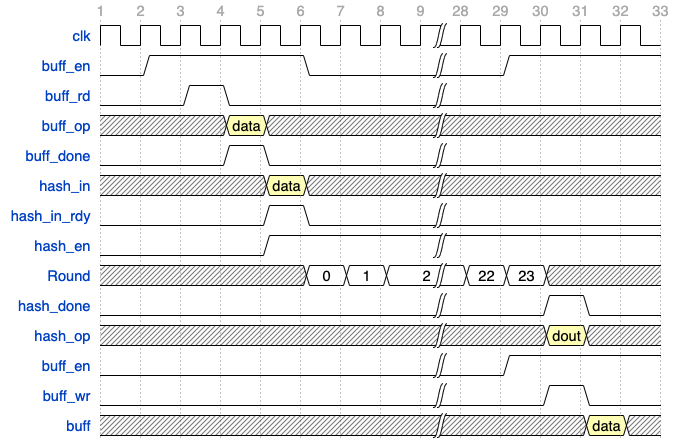}
\vspace{-6pt}
\caption{Timing Overhead of Instruction 17 \& 18}
\vspace{-15pt}
\label{fig:hb}
\end{figure}

\begin{figure}[!htb]
\centering
\vspace{-5pt}
\includegraphics[scale=0.35]{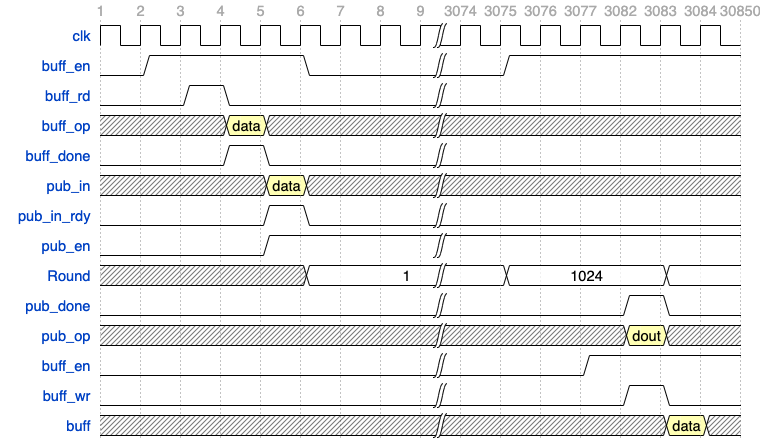}
\vspace{-8pt}
\caption{Timing Overhead of Instruction 19 \& 20}
\vspace{-15pt}
\label{fig:pub}
\end{figure}

The proposed architecture is implemented in Artix-7 (csg324-100t)FPGA using Vivado Tool. The additional hardwares added to the original architecture to protect the keys form major threats as stated in Table \ref{res} are $SC$, $buffer$ and the $CBI$. The signature checker consist of a KECCAK and RSA-1024 which cost 4188 and 31008 slices respectively. The base architecture without blockchain consumes 50k logic cells. The architecture with blockchain to protect severe key threats consumes 95k logic cells which is around 45\% of the total logic cells available in Artix-7 FPGA. As shown in Table \ref{res} articles \cite{aes:mccp}, \cite{hcrypt} and \cite{robupaul} proposed dedicated secret key memories which are completely isolated from processor area to prevent software threats but this architecture does not prevent incorrect use, insecure movement and non-destructions issues of keys. If any spoofing IPs or dishonest probes already exist inside the architectures of \cite{aes:mccp}, \cite{hcrypt} and \cite{robupaul} and try to read secret key stored in dedictaed key memory, the system will allow the key transaction to those malicious nodes. Though the dedicated key memories proposed in these article are physically isolated but can not prevent said incorrect and insecure movement of secret keys. Articles \cite{aes:mccp}, \cite{hcrypt} and \cite{robupaul} do not have any facility to investigate the key movement and to prevent non-destruction issue. Amazon Web Server \cite{aws} is one of the most recent software based key management architecture which can audit and investigate tracking of secret keys, but it cannot prevent the insecure movement and incorrect use of key. It does not have any signature checking facility on the key request.
In our architecture we have introduced dedicated secret key memory which is physically isolated from processor area. Any read write operation on this secret key memory named as $Master~Key~Memory$ (MKM) is blockchain based. We have observed the usual operations on MKM are related with $RNG$, $HASH$, and $AES$ block. The $Public$ and $Private$ keys of all these crypto IPs are already generated in offline by  an automatic script during the RTL development. The $pre~master~key$ is generated by $RNG$ which needs to be written in $MKM$  through buffer as shown in Table \ref{ins} instructions 2, 3, and 17 to 21. Instruction 2 writes random number generated by $RNG$ into buffer. The buffer also stores the public keys of $RNG$ (Source IP) and $MKM$ (Destination). The buffer includes timestamp form $TIMER$ IP, and stores all the current status of $Done$ and $Enable$ pin of available IPs in the proposed design. The instructions 17 to 21 generate signature of the write operation on $MKM$ by $RNG$. This signature will also be stored in $buffer$. This signature will be verified by the $Signature~Checker$ IP placed inside $Confidential~Area$ as stated in fig.\ref{fig:sys}. If the signature is matched then the $pre~master~key$ generated by $RNG$ will be written in $MKM$, otherwise the transaction will be discarded form the $buffer$. The other read write requests on $MKM$ are (ii)read $pre~master~key$ by $HASH$t to generate $master~key$ (Instruction 7 and 8)(ii)write $master keys$ in $MKM$ (Instruction 9 and 10). (iii)read $master~key$ form $MKM$ by $hash$ (Instruction 14 and 15)and (iv) read $master~key$ form $MKM$ by $AES$ (Instruction 11 and 13). All these read write operations follows Instruction 17 and 21 for authentic checking of the requestee. The verification process of $signature~checker$ IP using the $buffer$ data  with system status, timestamp, signature and the $pre~hash$ prevents incorrect use and insecure movement of keys. The $MKM$ also delete the keys which are already read to address non-destruction issue. The partitioned memory for secret keys prevents software attacks. The timing diagram of instruction 17, 18 and 19, 20 is shown in fig. \ref{fig:hb} and fig. \ref{fig:pub} respectively. Table \ref{tof} shows the trade off of latency and resource usage with security features. 

\begin{table}
	\setlength{\tabcolsep}{2.5pt}
	\vspace{-0.9em}    
	\caption{Summery of Trade-off}
	\label{tof}
	\resizebox{8.5cm}{!}{
		\begin{tabular}{|c|c|c|c|c|c|}\hline
&Partition &	Path  &		\multicolumn{2}{c|}{SC}&	Base \\ \cline{4-5} 
	&Memory &	Controller &	RSA	& KECCAK &	 Design\\ \hline
Slice &	0 &	76 &	31008 &	4188 &	50223\\ \hline
BRAM &	7.5 &	0 &	0 &	0 &	32\\ \hline
Clock &	2 &	2 &	3048 &	24 &	NA\\ \hline
latency &	20 ns &	10 ns &	86us &	67.2 ns &	NA\\ \hline
\shortstack{Feature\\ / role} &\shortstack{prevent\\SW attacks}&\shortstack{control crypto \\ instrcutions}&\multicolumn{2}{c|}{\shortstack{prevent insecure movment \& \\ incorrect use of keys; audit log}}&\shortstack{implements basic hw \\for secuirty processor}\\\hline

		\end{tabular}
	}
	\tiny{NA=Base desgn consists of diferent hardwares having different latency \& clock}
	\vspace{-15pt}
\end{table}

\vspace{-8pt}
\section{Conclusions}
\label{sec:con}
This article proposes a hardware blockchain with a partitioned and dedicated secret key memory which prevents most of the  software attacks, insecure key movement, incorrect use, non-destruction use secret key. Apart from this, if breach of keys occurs in the system, the hardware blockchain can investigate previous key transactions. To the best of our knowledge, the FPGA based blockchain to prevent threats stated in Table \ref{res} is never explored. The additional hardware adopted for blockchain is very nominal in-terms of resource usage and throughput.



%
\bibliographystyle{unsrt}  
\bibliography{IEEEexample}

\end{document}